\documentclass[doublecol]{epl2} 
\usepackage{amsmath}
\usepackage{amssymb}
\DeclareMathOperator{\sign}{sign}
\DeclareMathOperator{\Tr}{Tr}

\title{The picture of the Bianchi I model via gauge fixing in Loop Quantum Gravity.}
\shorttitle{The picture of the Bianchi I model...} %Insert here a short version of the title if it exceeds 70 characters

\author{F. Cianfrani\inst{1} \and A. Marchini\inst{1} \and G. Montani\inst{1,2,3}}
\shortauthor{F. Cianfrani \etal}

\institute{                    
  \inst{1} Dipartimento di Fisica, Universit\`a di Roma ``Sapienza'', Piazzale Aldo Moro 5, 00185 Roma, Italy.\\
  \inst{2} ENEA, Centro Ricerche Frascati, U.T. Fus. (Fus. Mag. Lab.), Via Enrico Fermi 45, 00044 Frascati, Roma, Italy.\\   
  \inst{3} INFN, Sezione Roma1, Piazzale Aldo Moro 5, 00185 Roma, Italy.
}
\abstract{The implications of the SU(2) gauge fixing associated with the choice of invariant triads in Loop Quantum Cosmology are discussed for a Bianchi I model. In particular, via the analysis of Dirac brackets, it is outlined how the holonomy-flux algebra coincides with the one of Loop Quantum Gravity if paths are parallel to fiducial vectors only. This way the quantization procedure for the Bianchi I model is performed by applying the techniques developed in Loop Quantum Gravity but restricting the admissible paths. Furthermore, the local character retained by the reduced variables provides a relic diffeomorphisms constraint, whose imposition implies homogeneity on a quantum level. The resulting picture for the fundamental spatial manifold is that of a cubical knot with attached SU(2) irreducible representations. The discretization of geometric operators is outlined and a new perspective for the super-Hamiltonian regularization in Loop Quantum Cosmology is proposed.}

\begin{document}

\maketitle

\section{INTRODUCTION}
Loop Quantum Gravity (LQG) \cite{librov, libthi, reviewQG} looks the most promising approach to the quantization of the gravitational field and it  stems from rewriting General Relativity as a gauge theory of the SU(2) group through the so-called Ashtekar-Barbero variables\cite{ashvar}. The quantum theory is defined in order to preserve background independence like in the classical theory. The most outstanding results are discrete spectra for geometric quantum operators in the kinematical Hilbert space, leading to a granular structure of the space-time at the Planck scale\cite{a&l1, a&l2}. Despite these results there are several technical difficulties in studying the dynamics and two main approaches have been developed in this sense: the Master Constraint Program\cite{MC} and Spinfoam models\cite{sf}. 

The investigation of LQG becomes easier by considering those cases in which the presence of proper symmetries simplifies the whole formulation. This is the case of Loop Quantum Cosmology (LQC)\cite{a&p}, where the techniques developed in LQG are applied to a homogeneous and, even, isotropic space. It has been shown that LQC models are free from the cosmological singularity\cite{boj1, boj2}, while recently the phenomenological implications on the spectrum of the cosmic microwave background radiation have been discussed\cite{b&c, grain}.

However, LQC does not represent the cosmological sector of LQG. For instance, the spectra of geometric operators in the kinematical Hilbert space of LQC is continuous and it becomes discrete only after the regularization of the Super-Hamiltonian operator. Such a regularization procedure involves the introduction of a discrete parameter $\bar\mu$, whose value is chosen in order to get the same minimum area eigenvalue as in LQG \cite{ash2006}. The whole information about the quantum space-time structure is thus contained into the fundamental length $\bar\mu$ and its value determines the physical properties of the scenario replacing the initial singularity. 

In order to give a theoretical explanation for $\bar\mu$, the foundation of LQC has been analyzed in \cite{CM,CMcr}, by retaining the SU(2) gauge structure in a cosmological framework. Geometric operators spectra were discrete, while the proper classical limit for the super-Hamiltonian was inferred if $\bar\mu^3$ coincided with the total number of verticies of the quantum graph underlying the continuous spatial picture. However, such a value for $\bar\mu$ did not provide a viable phenomenological scenario for the early Universe dynamics in the presence of a clock-like scalar field. 

Therefore, a deeper investigation on the relationship between LQC and LQG has started in \cite{CM1} for a Friedmann-Robertson-Walker (FRW) space-time. In that paper, the implications of the SU(2) gauge fixing associated with the restriction to invariant triads have been analyzed in details, by determining the associated Dirac brackets. The requirement to preserve the holonomy-flux algebra of the full theory led to choose a suitable class of restricted paths. Furthermore, reduced variables retained a local character and this feature implied that a subgroup of the diffeomorphisms group was preserved as gauge symmetry. The imposition of such a gauge invariance has been done via standard LQG techniques and it provided the implementation of homogeneity and isotropy on a quantum level. 

In this work we consider the homogeneous Bianchi I cosmological model. In particular, we will adopt the configuration space of LQG and we will analyze the implications of the SU(2) gauge-fixing leading to invariant triads. We will see how in order to preserve the holonomy-flux algebra of the full theory, the reduction to suitable paths must be implemented. Henceforth, the quantization techniques proper of LQG can be applied by considering such reduced paths only. Then, reduced diffeomorphisms will coincide with translations along some fiducial directions and the associated gauge invariance will imply homogeneity on a quantum level. Hence, we will outline how the action of reduced fluxes on functionals defined over reduced graphs will give geometric operators with discrete spectra. Such a discretization will emerge because of the compactness of the gauge group, but the obtained spectrum will be different from the one of LQG. As a consequence, the regularization of the Super-Hamiltonian constraint adopted in LQC will be questioned. Thereby, a regularization procedure will be discussed in which the parameter $\bar\mu$ is related with the number of vertices of the fundamental graph structure underlying the spatial manifold. 

\section{LOOP QUANTUM GRAVITY}
In the Holst formulation for gravity phase space can be described via $(A^i_a,E^b_j)$, where the former are the Ashtekar-Barbero connections and the latter are their conjugates momenta, the densitized triads. In terms of these variables the theory has the structure of a SU(2) gauge theory. A non-distributional Poisson algebra can be defined by considering smeared quantities, such as the holonomies of the connections along a path $\Gamma(s)$ and the fluxes of the triads across a surface $S$ with normal $n_a$, whose algebra reads (in units $8\pi G=\hbar=c=1$)   
\begin{align}\label{hf}
\left[E_i(S),h_{\Gamma}\right]_{\mathrm{PB}}=\;\;\;\;\;\;\;\;\;\;\;\;\;\;\;\;\;\;\;\;\;\;\;\;\;\;\;\;\;\;\;\;\;\;\;\;\;\;\;\;\;\;\;\;\;\;\;\;\;\;\;\;\;\;\;\;\;\;\;\;&\notag\\
=
\begin{cases}
 -i\gamma\sum_Ao^{\Gamma,S}(s_A)h_{\Gamma}^{0,s_A}\tau_ih_{\Gamma}^{s_A,1} & \Gamma\cap S=\{\Gamma^i(s_A)\},\\
0 & \mbox{otherwise}
\end{cases},
\end{align}
where
\begin{equation}
o^{\Gamma,S}(s_A)=\frac{n_a(s_A)\frac{\upd\Gamma^a}{\upd s}\left|_{s_A}\right.}{\left|n_b(s_A)\frac{\upd\Gamma^b}{\upd s}\left|_{s_A}\right.\right|}.
\label{o}\end{equation}

$\gamma$ being the Immirzi parameter, while $\tau_i$ denote the Hermitian SU(2) generators.

The quantum configuration space is defined as the set of the distributional connections $\bar{X}$, {\it i.e.} the set of all the homomorphism between the set of all smooth piecewise analytic paths of the spatial manifold and the SU(2) group. This space can be built via the projective limit for the space of general homomorphism $X_{l(\Gamma)}$ from a generic path $\Gamma$ to the gauge group. The paths $\Gamma$ are just labels and a crucial property is that they form a partially ordered set. One can then demonstrate that $\bar{X}$ is a compact Hausdorff space in the Tychonov topology, such that starting from the SU(2) Haar measure a Borel measure $\upd\mu$ for $\bar{X}$ can be given. The kinematical Hilbert space for the model is $\mathcal{L}^2(\bar{X},\upd\mu)$, {\it i.e.} the space of square integrable functions over $\bar{X}$ with respect to the measure $\upd\mu$. A self-adjoint operator corresponding to $E^a_i(x)$ can be defined via the holonomy-flux algebra (\ref{hf})\cite{self}. The states for the theory are invariant spin-networks. This way, the quantum space is endowed with a combinatorial structure, given by the edges and the vertices of the fundamental paths underlying the continuous picture. This structure, together with the compactness of the gauge group, is responsible for the discreteness of the geometric operators spectra\cite{a&l1, a&l2}.
\section{LOOP QUANTUM COSMOLOGY}
Anisotropic models are expected to be relevant in the early phase of the Universe, approaching the cosmological singularity\cite{revcosm}. The Bianchi I model is the simplest homogeneous and anisotropic model and it describes a space with a null 3-curvature whose metric is invariant under translations along the three spatial directions in Cartesian coordinates. The line element of the model is described by three dynamic variables, which coincide with the three scale factors $a_a$, {\it i.e.} 
\begin{equation}
\label{dsBI}
\upd s^2=-\upd t^2+a_1^2(t)\upd x_1^2+a_2^2(t)\upd x_2^2+a_3^2(t)\upd x_3^2.
\end{equation}

We define also the fiducial metric ${}^0h_{ab}$ as the time-independent part of the spatial metric, whose line element is
\begin{equation}
 \label{fid}
{}^0\upd s^2=\upd x_1^2+\upd x_2^2+\upd x_3^2.
\end{equation}

By exploiting the symmetries of the model, one can define invariant connections\cite{b&k} and the densitized triads as
\begin{equation}\label{gf}
A^i_a=c^i{}^0\!e^i_a,\qquad E^a_i=p_i{}^0\!e{}^0\!e^a_i,
\end{equation} 
where $p_i=|\epsilon_{ijk}a_ja_k|\sign(a_i)$ and $c^i=\gamma\dot{a}_i$, while ${}^0\!e^a_i$ and ${}^0\!e^i_a$ denote the fiducial 1-forms and vectors, respectively, and ${}^0\!e=\det({}^0\!e^i_a)$. The reduced phase space is then described by the conjugate variables $(c^i,p_j)$, while the supermomentum and the Gauss constraint vanish identically. Holonomies can be defined along straight paths parallel to the fiducial triad ${}^0\!e^i_a$ and they read (the index $i$ is not summed)
\begin{equation}\label{hol}
 h_i=\exp(i\mu_i c^i\tau_i)%=\mathbb{I}_2\cos\left(\frac{\mu_a c^a}{2}\right)+2{}^j\!\tau_a\sin\left(\frac{\mu_a c^a}{2}\right)
,
\end{equation}
where $\mu_i$ is the edge length along the direction ${}^0\!e^i_a$. The Poisson brackets turn out to be
\begin{equation}
[p_i,h_j]_{\mathrm{PB}}=-i\frac{\gamma\mu_b}{V_0}h_j\tau_j\delta_{ij},\label{pb}
\end{equation}
$V_0$ being the volume in the fiducial metric. The configuration space of LQC is parametrized by the matrix elements of the holonomies' tensor product $h_1\otimes h_2\otimes h_3$, which are given by the quasi-periodic functions
\begin{equation}
 N_{\vec{\mu}}=N_{\mu_1}N_{\mu_2}N_{\mu_3}=\exp\left[\frac{i}{2}\left(\mu_1c^1+\mu_2c^2+\mu_3c^3\right)\right].
\end{equation}
The Poisson brackets $[p_a,N_{\vec{\mu}}]_{\mathrm{PB}}$ plays for LQC the same role as the holonomy-flux algebra for LQG and the space of distributional connections is the Bohr compactification of the real line $\textbf{R}_{\mathrm{Bohr}}$ for each direction \cite{CDW1,BI1,BI2}. The kinematical Hilbert space is $\textsc{H}=\mathcal{L}^2(\textbf{R}^3_{\mathrm{Bohr}},\upd\vec{\mu})$ whose scalar product can be inferred from $\langle N_{\vec{\mu}}|N_{\vec{\mu}'}\rangle=\delta_{\vec{\mu}\vec{\mu}'}=\delta_{\mu^{}_1\mu_1'}\delta_{\mu^{}_2\mu_2'}\delta_{\mu^{}_3\mu_3'}$. One can also define self-adjoint momentum operators from the Poisson brackets (\ref{pb}), and they act on quasi-periodic functions as follows 
\begin{equation}\label{p}
% \hat{p}_ah_b=-i\frac{\upd h_b}{\upd c^a}=\frac{\gamma l^2_\mathrm{P}\mu_b}{V_0}h_b{}^j\!\tau_b\delta_{ab}, \qquad 
\hat{p}_iN_{\vec{\mu}}=\frac{\gamma\mu_i}{V_0}N_{\vec{\mu}}.
\end{equation}
%$l_\mathrm{P}$ being the Planck length. 
The expression for the Super-Hamiltonian of the model in a proper factor ordering \cite{CDW1} reads
\begin{align}\label{HLQC}
  \mathcal{H}&=-\frac{V_0}{\gamma^2}\left[\frac{\hat{p}_1\hat{p}_2}{\sqrt{\hat{p}_1\hat{p}_2\hat{p}_3}}\frac{\hat{\sin}(\bar{\mu}_1c^1)\hat{\sin}(\bar{\mu}_2c^2)}{\bar{\mu}_1\bar{\mu}_2}+\right.\notag\\ &\left.+\frac{\hat{p}_1\hat{p}_3}{\sqrt{\hat{p}_1\hat{p}_2\hat{p}_3}}\frac{\hat{\sin}(\bar{\mu}_1c^1)\hat{\sin}(\bar{\mu}_3c^3)}{\bar{\mu}_1\bar{\mu}_3}\right.+\notag\\
 &\left.+\frac{\hat{p}_2\hat{p}_3}{\sqrt{\hat{p}_1\hat{p}_2\hat{p}_3}}\frac{\hat{\sin}(\bar{\mu}_2c^2)\hat{\sin}(\bar{\mu}_3c^3)}{\bar{\mu}_2\bar{\mu}_3}\right].
\end{align}

$\bar{\mu}_i$ for $(i=1,2,3)$ in the expression above denote the lengths of the edges along which holonomies are evaluated and they play the role of regulators in the definition of the quantum Super-Hamiltonian. However, $\bar\mu_i$ cannot be removed because the limit for $\bar{\mu}_i\to0$ of $\mathcal{H}$ does not exist. Hence, $\mathcal{H}$ is evaluated at a fixed value for $\bar{\mu}_i$. This value has been chosen in \cite{CDW1} such that the area operator in LQC along each direction (whose definition involves $\hat{p}_i$) has the same minimum eigenvalue as the area operator in LQG (see also \cite{BI1} where a different prescription is adopted for $\bar\mu_i$). At the end along each direction the initial singularity is avoided as in the isotropic case.  

\section{GAUGE-FIXED LOOP QUANTUM GRAVITY}
On a classical level, we are free to rotate the triads without changing the physical results of the theory because of SU(2) gauge invariance. Hence,  the restriction to invariant connections and triads (\ref{gf}) is actually a gauge-fixing of the SU(2) symmetry. In particular, the generic expression for $E^a_i$ can be obtained from (\ref{gf}) by a rotation, {\it i.e.}
\begin{equation}
E^a_i=\sum_{j}p_j\Lambda^j_i {}^0\!e^a_j, 
\end{equation}     
$\Lambda^j_i$ being a generic SO(3) matrix, which is arbitrary as soon as gauge invariance is preserved. The condition $\Lambda^j_i=\delta^j_i$ is fixed by 
\begin{equation}
\chi_i=\epsilon_{ij}^{\phantom{12}k}{}^0\!e^j_a E^a_k=0.\label{chi}
\end{equation}
Indeed, the relation above is solved by 
\begin{equation}
E^a_i=p_i(x,t){}^0\!e^a_i,\label{pa}
\end{equation}
where we have three different $p_i$ which are functions of all space-time variables. The condition (\ref{chi}) completely fixes SU(2) gauge freedom. This can be seen computing the Poisson brackets between $\chi_i$ and the Gauss constraint, \textit{i.e.}
\begin{equation}
G_i=\partial_aE^a_i+\gamma\epsilon_{ij}{}^kA^j_aE^a_k,
\end{equation}
which provide the following non-degenerate result
\begin{align}\label{GChi1}
 \left[G_i(x),\chi_j(y)\right]_{\mathrm{PB}}&\approx-\gamma{}^0e\delta_{ij}(p_1+p_2+p_3-p_b)\delta^{(3)}(x-y).
\end{align}
When the gauge conditions (\ref{chi}) is fixed the set of constraints becomes second class. The quantization of such a system can be done by working in reduced phase space (this is what generally is done in LQC) or by describing the constrain surface with the unreduced variables using the Dirac brackets instead of Poisson ones. In this work we follow the second way to infer the proper correspondence between reduced and unreduced coordinates. The Dirac brackets between the connections read
\begin{align}\label{DBC}
&\left[A^i_a(x),A^j_b(y)\right]_{\mathrm{DB}}=\left(K\left(A^i_b(x){}^0\!E^j_a(y)+\right.\right. \notag  \\ &\left.\left.-{}^0\!E^i_b(x)A^j_a(y)+2\delta^{ij}\delta_{kl}{}^0\!E^k_{[b}(y)A^l_{a]}(x)\right)+\left(K_i+\right.\right.\notag \\ &\left.\left.+K_j\right)\left({}^0\!E^i_b(x)A^j_a(y)-A^i_b(x){}^0\!E^j_a(y)\right)+2\left(K_i+\right.\right.\notag \\ &\left.\left.+K_k\right)\delta^{ij}\delta_{kl}{}^0\!E^k_{[a}(y)A^l_{b]}(x)\right)\delta^{(3)}\left(x-y\right)-\frac{\epsilon^{ij}{}_k}{\gamma}\left(K_i+\right.\notag \\ &\left.+K_j\right)\left({}^0\!E^k_b(x)\frac{\partial\delta^{(3)}(x-y)}{\partial y^a}+{}^0\!E^k_a(y)\frac{\partial\delta^{(3)}(x-y)}{\partial x^b}\right), 
\end{align}
where
\begin{equation}
  K_i=\frac{1}{p_1+p_2+p_3-p_i}, \qquad K=\sum_{i=1}^3K_i.
\end{equation}
We can see from (\ref{DBC}) that, after the gauge-fixing, the connections do not commute because their components are not all independent. The remaining brackets are
\begin{align}
 &\left[A^i_a(x),E^b_j(y)\right]_{\mathrm{DB}}=\left(\delta^i_j\delta^b_a+K\left({}^0\!E_{aj}(y)E^{bi}(x)+\right.\right.\notag \\ &\left.\left.-\delta^i_j{}^0\!E^m_a(y)E^b_m(x)\right)-\left(K_i+K_j\right){}^0\!E_{aj}(y)E^{bi}(x)+\right.\notag \\ &\left.+\delta^i_j(K_j+K_m){}^0\!E^m_a(y)E^b_m(x)\right)\delta^{(3)}\left(x-y\right),
\end{align}
\begin{align}
 \left[E^a_i(x),E^b_j(y)\right]_{\mathrm{DB}}=0.
\end{align}
We can select the following independent components of the connections on the constraint hypersurfaces
\begin{equation}\label{cb}
 c^1=A^1_a{}^0\!e^a_1, \quad c^2=A^2_a{}^0\!e^a_2, \quad c^3=A^3_a{}^0\!e^a_3,
\end{equation}

which coincide with reduced variables. Their Dirac brackets are given by
\begin{align}
  &\left[c^i(x,t),c^j(y,t)\right]_{\mathrm{DB}}=0,\\
  &\left[c^i(x,t), E_j^b(y)\right]_{\mathrm{DB}}={}^0\!e_j^{b}(y)\delta^{i}_j\delta^{(3)}(x-y).\label{par-hol}
\end{align}
We can rewrite a generic holonomy by using the relations (\ref{cb}) as
\begin{align}
  h_{\Gamma}=e^{i\int_{\Gamma}A^i_a\frac{\upd\Gamma^a}{\upd s}\upd s\tau_i}=e^{i\sum_{i=1}^3\int_{\Gamma}c^i\left(x(s)\right){}^0\!e^i_a\frac{\upd\Gamma^a}{\upd s}\upd s\tau_i},
\end{align}
and we can infer the action of the unreduced momenta after the gauge-fixing through (\ref{par-hol}). The holonomy-flux algebra after the gauge-fixing is
\begin{align}\label{hf-gf}
\left[E_i(S),h_{\Gamma}\right]_{\mathrm{DB}}=\;\;\;\;\;\;\;\;\;\;\;\;\;\;\;\;\;\;\;\;\;\;\;\;\;\;\;\;\;\;\;\;\;\;\;\;\;\;\;\;\;\;\;\;\;\;\;\;\;\;\;\;\;\;\;\;\;\;\;\;&\notag\\
=
\begin{cases}
 -i\gamma\sum_A\widetilde{o}_i^{\Gamma,S}(s_A)h_{\Gamma}^{0,s_A}\tau_ih_{\Gamma}^{s_A,1} & \Gamma\cap S=\{\Gamma^a(s_A)\},\\
\int C_i h_\Gamma(0,s)\tau_i h_\Gamma(s,1)\upd s & \Gamma\subset S,\\
0 & \mbox{otherwise}
\end{cases},
\end{align}
where
\begin{equation}
\widetilde{o}^{\Gamma,S}_i(s_A)=\displaystyle\frac{n_a(s_A){}^0\!e^a_i(s_A) (\upd\Gamma^b/\upd s)|_{s_A}{}^0\!e_b^i(s_A)}{|n_c(s_A) (\upd\Gamma^c/\upd s)|_{s_A}|},\label{oo}
\end{equation}
while $C_i$ is a factor, which would need to be regularized. The index $i$ in the expression (\ref{oo}) is not summed. In what follows, repeated gauge indexes will not be summed unless explicitly mentioned. 

The expression (\ref{hf-gf}) gives the holonomy-flux algebra as soon as the gauge-fixing condition (\ref{chi}) holds. In this case it is not possible to represent the action of fluxes in terms of left-invariant vector fields as in LQG. Furthermore, there is also a diverging contribution in the case when $\Gamma$ belongs to $S$. 

These problems can be avoided if we choose to consider, among all the possible paths, only those ones $\Gamma^i$ tangent to fiducial vectors, \textit{i.e.} $\frac{\upd(\Gamma^i)^a}{\upd s}\propto{}^0\!e^a_i$. We denote this paths as reduced paths. When holonomies are evaluated on reduced paths, the factor $\widetilde{o}^{\Gamma,S}_i(s_A)$ is equal to the expression (\ref{o}) and the divergent term that appears when $\Gamma\subset S$ is tamed because $C_i=0$. In other words, the holonomy-flux algebra after the gauge-fixing is still the one of the LQG if we consider only the holonomies associated with the paths parallel to the fiducial vectors, so we are forcing to choose only the following three classes of holonomies
\begin{equation}
   h_{\Gamma^i}=e^{i\int_{\Gamma^i}c^i(s)\upd s\tau_i},
\end{equation}

The restriction to reduced paths is a standard tool in LQC. Here, we can motivate this choice with the requirement to reproduce the holonomy-flux algebra of LQG, which will allow us to define the action of variables associated with fluxes as essentially self-adjoint operators.

%In Section 2 we have described how to construct the kinematical Hilbert space for LQG. 
Let us now defined the quantum configuration space associated with holonomies along reduced paths. The main point is that reduced paths still form a partially ordered set. Hence, the definition of the space of reduced distributional connections $\bar{X}_\mathrm{C}$ for the Bianchi I model can be performed as in LQG. $\bar{X}_{\mathrm{C}}$, as $\bar{X}$, is a compact Hausdorff space and we can define on it the Ashtekar-Lewandowski measure. One can introduce the kinematical Hilbert space $\mathcal{L}^2(\bar{X}_{\mathrm{C}},\upd\mu)$ and basis vector on this space are invariant spin networks defined on the reduced paths. Therefore, the quantum spatial manifold underlying the Bianchi I model takes the form of a graph with cubic topology.   Proper essentially self-adjoint flux operators are inferred from (\ref{hf-gf}) and they act as follows
 \begin{align}
 &\hat{E}_i(S)h_{\Gamma^j}=\nonumber\\ &=
\begin{cases}
 \gamma\sum_A\delta_i^jo^{\Gamma^j,S}(s_A)h_{\Gamma^j}^{0,s_A}\tau_ih_{\Gamma^j}^{s_A,1} & \Gamma^j\cap S=\left\{(\Gamma^j)^b(s_A)\right\},\\
0 & \mathrm{otherwise}
\end{cases}\label{ehat}
\end{align}
The relations above demonstrate that the flux operators associated with $E^a_i$ are oriented along the fiducial vectors. Only by taking properly into account the gauge-fixing (\ref{chi}) the resulting quantum geometric structure reflects the one proper of fiducial vectors. When the flux operator is well-defined, we can introduce the area operator for a surface $S_j$, whose normal coincide with ${}^0\!e^j_a$ and which intersects $\Gamma^i$ in $s_A$. This operator needs to be regularized as in \cite{a&l1} and the final expression reads
\begin{equation}
\hat{A}[S_j]h_{\Gamma^i}=\delta^i_j|\gamma|\sum_A n_a(s_A){}^0\!E^a_i(s_A) h_{\Gamma^i}|\tau_i|.%%%legame area LQC!!!!!!!
\end{equation}

To build up the space of the diffeomorphisms invariant states the action of the corresponding constraints must be discussed. In fact, the gauge condition (\ref{chi}) imposes the restriction $E^a_i=p_i(x,t){}^0\!e^a_i$ and $A^i_a=c^i(x,t){}^0\!e^i_a$, for which the supermomentum does not identically vanish. This is the case because the reduced variables $c_a, p_b$ retain a dependence on spatial coordinates. Hence, on a quantum level some restrictions have to be imposed in order to find out the physical Hilbert space. In this respect let us now consider  the generator of 3-diffeomorphisms, {\it i.e.}
\begin{eqnarray}
D[\vec{\xi}]=\sum_i\int [\xi^a E^b_i\partial_aA^i_b-\xi^a \partial_b(A^i_aE^b_i)]d^3x%=\nonumber\\=\int \sum_a[\xi^ip_a\partial_ic^a-\xi^i\partial_j(p_ac^a{}^0\!e^a_i {}^0\!e^j_a)]d^3x
,
\end{eqnarray}
$\xi^a$ being arbitrary parameters, whose action in reduced phase-space reads
\begin{equation}
[D[\vec{\xi}],c^j(x)]_{\mathrm{DB}}=[\xi^a\partial_ac^j+c^j{}^0\!e^j_a {}^0\!e^b_j \partial_b\xi^a]|_x.
\end{equation}

The induced finite transformation on reduced holonomies is given by
\begin{equation}
[D[\vec\xi],h_{\Gamma^i}]_{\mathrm{DB}}=h_{\phi(\Gamma^i)}-h_{\Gamma^i},
\end{equation}

where the path $\phi(\Gamma^i)$ is obtained from $\Gamma^i$ by the translation $x^i\rightarrow x^i-\xi^a{}^0\!e^i_a$, $x^i$ being the coordinate along ${}^0\!e^i_a$, {\it i.e.} $x^i=x^a{}^0\!e^i_a$. Henceforth, the full 3-diffeomorphisms invariance reduces to the invariance under translations along fiducial vectors.

Translational invariant states can be defined as in \cite{CM1} by working with reduced knots,  \textit{i.e.} knot classes built as equivalence classes of reduced paths under translations. The resulting spatial manifold exhibits a fundamental homogeneity along each direction. This is due to the fact that the whole diffeomorphisms group has been reduced to the invariance under translations along each fiducial direction. The description of a cosmological space in terms of a knot with cubic topology and a fixed quantum state attached to each edge will allow us to discuss a self-consistent cosmological implementation of LQG and to investigate the foundation of LQC.

\section{LQC OPERATORS}
In this paragraph we discuss the link existing between reduced variables and the operators of LQC. 

Let us now consider reduced holonomies (\ref{hol}) and fluxes across the surfaces $S_i$, \textit{i.e.}
\begin{align}\label{fluxBI}
 E_i(S_i)&=p_i\Delta_i, & \Delta_i&=\int_{S_i}{}^0\!e{}^0\!e^a_in_a\upd u\upd v=\int_{S_i}{}^0\!e\upd u\upd v,
\end{align} 
where $\Delta_i$ is the measure of the surface $S_i$ in the fiducial metric. From the relation (\ref{ehat}), the action of momenta operators on reduced holonomies is given by
\begin{equation}
 \hat{E}_i(S_i)h_j=\gamma h_j\tau_j\delta^i_j\sign(\Delta_i\mu_j).\label{2}
\end{equation}
%and from (\ref{fluxBI}) one can define the operator associated with $p_a$ in LQG as follows
%\begin{equation} \hat{p}_a\Delta_ah_b=\gamma h_b{}^j\!\tau_b\delta^a_b\sign(\Delta_a\mu_a).\label{3}\end{equation}

%The operator above provides a non-vanishing result only for holonomies along the same direction as $S_a$, {\it i.e.} $a=b$. Hence, as soon as the tensor product of the holonomies is concerned, one finds (for instance for $a=1$) \begin{equation}\label{p_1}\hat{p}_1\left(h_1\otimes h_2\otimes h_3\right)=\left(\frac{\gamma l^2_P\sign(\Delta_1\mu_1)}{\Delta_1}h_1{}^j\!\tau_1\right)\otimes h_2\otimes h_3.\end{equation}

The relations (\ref{fluxBI}), (\ref{2}) and (\ref{p}) are consistent if
\begin{equation}
 |\Delta_i{\mu_i}|=V_0.\label{d}
\end{equation}

The condition (\ref{d}) fixes a fundamental duality between the length of edges along which holonomies are evaluated and the area of those surfaces across which fluxes are defined. The elements of LQC Hilbert space can be inferred by a tracing over the SU(2) group indexes, {\it i.e.} 
\begin{equation}
\Tr h_i=\sum_{n=0}^{j_i-\theta}\cos(\mu_ic^i(n+\theta))+(1-2\theta).
\end{equation}
where $j_i$ is the spin quantum number of the SU(2) representation labeling the edge $e_i$, while $\theta$ is equal to $1/2$ and $0$ for $j_i$ half-integer and integer, respectively.

As for the mapping of momenta operators, one finds
\begin{align}
 &\Tr E_1(S_1)h_1\otimes h_2\otimes h_3=2|\Delta_1|\hat{p}_1\sum_{n=0}^{j_1-\theta}\cos(\mu_1c^1(n+\theta))+\notag\\
&+(1-2\theta)\otimes h_2\otimes h_3=\left(\gamma\Tr\left(h_1{}^{j_1}\!\tau_1\right)\right)\otimes h_2\otimes h_3=\notag\\
 &=\left(\gamma \sum_{n=0}^{j_1-\theta}i2(n+\theta)\sin(\mu_1c^1(n+\theta))\right)\otimes h_2\otimes h_3,\label{tr}
\end{align}
From the relations (\ref{tr}) and (\ref{d}), we obtain that in the space of quasi-periodic functions $\hat{p}_i$ is a derivative operator whose action reads for $i=1$
\begin{equation}
 \hat{p}_1\left(e^{i\tilde{\mu}_1c^1}\otimes e^{i\tilde{\mu}_2c^2}\otimes e^{i\tilde{\mu}_3c^3}\right)=\hat{p}_1e^{i\tilde{\vec{\mu}}\cdot\vec{c}}=\frac{\gamma}{V_0}\tilde{\mu}_1e^{i\tilde{\vec{\mu}}\cdot\vec{c}},
\end{equation}

with $\tilde{\mu}_i=m_i\mu_i$, $m_i$ being the magnetic spin quantum number. %Hence, the spectrum of $\hat{p}_a$ is discrete and this fact marks a difference with standard LQC, where the operator has a continuum spectrum. 
Now is possible to define a regularized area operator with a discrete spectrum in the kinematical Hilbert space because % in a cosmological contest without importing \textit{ad hoc} the discretization from LQG fundamental properties
\begin{equation}
 \hat{A}[S_1]e^{i\tilde{\vec{\mu}}\cdot\vec{c}}=\sqrt{\Delta_1^2\hat{p}^2_1}e^{i\tilde{\vec{\mu}}\cdot\vec{c}}=\gamma \left|\frac{\tilde{\mu}_1}{\mu_1}\right|e^{i\tilde{\vec{\mu}}\cdot\vec{c}}=\gamma |m_1|e^{i\tilde{\vec{\mu}}\cdot\vec{c}}.
\end{equation}
 
The spectrum of the area operator is determined by magnetic quantum numbers $m_i$ only, thus it is discrete in view of the gauge group compactness. Indeed, the spectrum does not coincide with one of LQG. Therefore, the procedure adopted in LQC to infer the polymer-like $\bar\mu_i$ (see for example\cite{CDW}) cannot be justified, because the spectrum of the area operator does not contain any dependence from $\mu_i$. The existence of a minimum value for $\mu_i$ is thus not a consequence of the kinematical properties of geometric operators in LQG. This shortcoming leaves open the question about the proper implementation of the dynamic. The Super-Hamiltonian in LQG is \cite{QSD}
\begin{equation}
 \mathcal{H}=-\frac{2}{\gamma^3}\sum_\mathrm{v}\mathcal{H}_\mathrm{v},
\end{equation}

where the sum is extended over all vertices of the graph on which the Super-Hamiltoniana acts and $\mathcal{H}_\mathrm{v}$ has the following form
\begin{equation}\label{Hv}
\mathcal{H}_\mathrm{v} =-\epsilon^{ijk}\Tr\left[h(s_{ij})h(s_k)\left[V,h^{-1}(s_k)\right]\right].
\end{equation}

$s_ {ij}$ denotes the rectangle having $v$ as one of its vertices and whose sides are contained into the edges emanating from $v$ along the $i$ and $j$ directions (the use of rectangular instead of triangular loops allows to have a consistent regularization procedure after the restriction to reduced paths (\ref{hol})). $s_k$ is merely an edge along the $k$ direction. All holonomies in the expression (\ref{Hv}) are in the fundamental representation and $V$ is the volume operator in the full space. The restriction to Bianchi I model gives the following volume operator
\begin{equation}
 V=V_0\hat{p}_1^{\frac{1}{2}}\hat{p}_2^{\frac{1}{2}}\hat{p}_3^{\frac{1}{2}},
\end{equation}

such that the commutator inside the expression (\ref{Hv}) acts on $h_1$ as 
\begin{equation}
 V_0\left[\hat{p}_1^{\frac{1}{2}}\hat{p}_2^{\frac{1}{2}}\hat{p}_3^{\frac{1}{2}},h_1\right]h_1=\gamma \bar{\mu}_1\frac{\hat{p}_2\hat{p}_3}{\sqrt{\hat{p}_1\hat{p}_2\hat{p}_3}}{}^{\frac{1}{2}}\!\tau_1h_1.
\end{equation}

The factor $\bar{\mu}_1$ represents the value of $\mu_1$ at which the regularization of the Super-Hamiltonian takes place. Assuming that each vertex gives the same contribution we obtain for the Super-Hamiltonian
\begin{align}\label{H_om}
 \mathcal{H}&=-\frac{N_\mathrm{v}}{\gamma^2
}\left[\frac{\bar{\mu}_1\bar{\mu}_2\bar{\mu}_3}{\bar{\mu}_1\bar{\mu}_2}\frac{\hat{p}_1\hat{p}_2}{\sqrt{\hat{p}_1\hat{p}_2\hat{p}_3}}\hat{\sin}(\bar{\mu}_1c^1)\hat{\sin}(\bar{\mu}_2c^2)+\right.\notag \\ &\left.+\frac{\bar{\mu}_1\bar{\mu}_2\bar{\mu}_3}{\bar{\mu}_1\bar{\mu}_3}\frac{\hat{p}_1\hat{p}_3}{\sqrt{\hat{p}_1\hat{p}_2\hat{p}_3}}\hat{\sin}(\bar{\mu}_1c^1)\hat{\sin}(\bar{\mu}_3c^3)\right.+\notag\\
 &+\left.\frac{\bar{\mu}_1\bar{\mu}_2\bar{\mu}_3}{\bar{\mu}_2\bar{\mu}_3}\frac{\hat{p}_2\hat{p}_3}{\sqrt{\hat{p}_1\hat{p}_2\hat{p}_3}}\hat{\sin}(\bar{\mu}_2c^2)\hat{\sin}(\bar{\mu}_3c^3)\right],
\end{align}

$N_\mathrm{v}$ being the total number of vertices. We can obtain from (\ref{H_om}) the Super-Hamiltonian (\ref{HLQC}) (which reproduces the proper classical limit \cite{CDW1}) by fixing
\begin{equation}\label{nver}
 \bar{\mu}_1\bar{\mu}_2\bar{\mu}_3=\frac{V_0}{N_\mathrm{v}}.
\end{equation}

Therefore, as in the case of homogeneous and isotropic model \cite{CM,CMcr}, $\bar\mu_i$ is determined by the total number of vertices of the graph underlying the continuous spatial structure.  

\section{CONCLUSIONS}
In this work, the foundation of LQC has been discussed investigating, in particular, the implications of the SU(2) gauge fixing by which the invariant triads of a Bianchi I model could be chosen in LQG phase space. We inferred the expression of the Dirac brackets, through which we could analyze the holonomy-flux algebra after the gauge fixing. We found that such an algebra coincided with the one of LQG for holonomies whose edges were along fiducial vectors only. Therefore, by restricting admissible paths it was possible to apply LQG quantization procedure to define the kinematical Hilbert space of the Bianchi I model. Henceforth, we inferred a fundamental combinatorial description of such an anisotropic cosmological model in terms of paths with a cubic topology. Furthermore, the presence of a relic diffeomorphisms symmetry in reduced phase space (associated with translations along fiducial vectors) could be implemented by defining physical states over reduced knots. This procedure allowed us to implement homogeneity on a quantum level, thus extending to the Bianchi I case the result obtained for the isotropic model in \cite{CM1}, {\it i.e.} the description of an homogeneous space-time in terms of a fundamental cubic and homogeneous knot-classes. This issues constitutes the main result of this analysis, because it opens up the possibility to discuss the dynamics of minisuperspace models by applying directly Thiemann's regularization procedure \cite{QSD} for the super-Hamiltonian operator.    

Then, we discussed the relationship of our reduced variables with LQC ones. In particular, we could establish a map between LQG and LQC Hilbert spaces through the trace over SU(2) indexes. Moreover, the consistency between the self-adjoint momenta operators in the two formulations required to fix a duality relation between the length of edges along which holonomies were evaluated and the area of surfaces across which fluxes were defined. As a consequence, the area operator retains a discrete spectrum and it does not depend on the length of edges. This point conflicts with the regularization procedure of the super-Hamiltonian adopted in LQC. Nevertheless, we emphasized how in order to reproduce the proper semi-classical limit, the parameters $\bar\mu_i$ at which the super-Hamiltonian had to be evaluated should be related with the total number of vertices of the fundamental cubic path underlying the continuous spatial picture. 

This result outlines how the proposed analysis on the cosmological sector of LQG can help in understanding the origin of the parameters $\bar\mu_i$ which enters the regularization of the super-Hamiltonian operator in LQC.

\end{document}